\documentclass[preprint]{aastex}

\newcommand{\Mpc}{\hbox{$\rm\thinspace Mpc$}}

\newcommand{\km}{{\rm\thinspace km}}
\newcommand{\cm}{{\rm\thinspace cm}}
\newcommand{\s}{{\rm\thinspace s}}
\newcommand{\kms}{\hbox{$\km\s^{-1}\,$}}
\newcommand{\degsq}{\hbox{$\deg^{2}\,$}}
\newcommand{\bj}{\hbox{$b_J$}}
\newcommand{\msun}{\hbox{${\rm\thinspace M}_{\sun}$}}
\newcommand{\lsun}{\hbox{${\rm\thinspace L}_{\sun}$}}

\slugcomment{Accepted 2000 December 19 for publication in ApJ Letters}
\shorttitle{Fornax substructure and dynamics}
\shortauthors{Drinkwater et al.}

\begin{document}

\title{Substructure and dynamics of the Fornax Cluster}

\author{Michael J.\   Drinkwater}
\affil{School of Physics, University of Melbourne, Victoria
3010, Australia}
\email{mjdrin@unimelb.edu.au}

\author{Michael D.\ Gregg}
\affil{Physics Dept., U.C. Davis, and IGPP, Lawrence Livermore
National Laboratory, L-413, Livermore, CA 94550, USA}
\email{gregg@igpp.ucllnl.org}

\and

\author{Matthew Colless}
\affil{Research School
of Astronomy \& Astrophysics, The Australian National University,
Weston Creek, ACT 2611, Australia}
\email{colless@mso.anu.edu.au}

\begin{abstract}
We present the first dynamical analysis of a galaxy cluster to include
a large fraction of dwarf galaxies.  Our sample of 108 Fornax Cluster
members measured with the UK Schmidt Telescope FLAIR-II spectrograph
contains 55 dwarf galaxies ($15.5>\bj>18.0$ or $-16>M_B>-13.5$).
H$\alpha$ emission shows that $36\pm8\%$ of the dwarfs are
star-forming, twice the fraction implied by morphological
classifications.  The total sample has a mean velocity of
$1493\pm36\kms$ and a velocity dispersion of $374\pm26\kms$.  The
dwarf galaxies form a distinct population: their velocity dispersion
($429\pm41\kms$) is larger than that of the giants ($308\pm30\kms$) at
the 98\% confidence level. This suggests that the dwarf population is
dominated by infalling objects whereas the giants are virialized.

The Fornax system has two components; the main Fornax Cluster centered
on NGC~1399 with $\overline{cz}=1478\kms$ and $\sigma_{cz}=370\kms$,
and a subcluster centered 3 degrees to the south-west including
NGC~1316 with $\overline{cz}=1583\kms$ and $\sigma_{cz}=377\kms$. This
partition is preferred over a single cluster at the 99\% confidence
level.  The subcluster, a site of intense star formation, is bound to
Fornax and probably infalling towards the cluster core for the first
time.  We discuss the implications of this substructure for distance
estimates of the Fornax Cluster.  We determine the cluster mass
profile using the method of \citet{dia1999}, which does not assume a
virialized sample.  The mass within a projected radius of 1.4\Mpc\ is
$(7\pm2)\times10^{13}\msun$, and the mass-to-light ratio is $300\pm100
\msun/\lsun$.  The mass is consistent with values derived from the
projected mass virial estimator and X-ray measurements at smaller
radii.
\end{abstract}


\section{Introduction}
\label{sec_intro}

As the largest gravitationally bound systems in the Universe, clusters
of galaxies offer unique constraints on the formation of large scale
structure: models of galaxy infall onto clusters predict the velocity
distributions of the infalling galaxies in terms of the density
enhancement of the cluster and the cosmological mass density
$\Omega_0$ \citep{reg1989}.  On smaller scales, clusters also provide
important clues for theories of galaxy formation and evolution: the
broader distribution of late-type (star-forming) galaxies compared to
early-types in both poor clusters \citep[e.g.\ Virgo,][]{huc1985} and
rich clusters \citep[e.g.\ Coma,][]{col1996} is generally interpreted
as evidence that the late-type galaxies are falling into the clusters
for the first time, whereas the early types are an older, virialized
population.  These observations are based on luminous cluster galaxies
($M_B<-16$), even though they are greatly outnumbered by dwarf
galaxies.  Studies of the dwarf populations are limited to small
samples in nearby clusters and groups, such as the Fornax Cluster
\citep{hel1994} and the Sculptor and Centaurus~A groups
\citep{cot1997}.  In these studies, the dwarfs had more extended
velocity and spatial distributions than the giants, but not at
statistically significant levels.  It is timely to investigate the
behavior of dwarfs in clusters using larger samples, as numerical
simulations of cluster formation can now resolve the formation of dark
matter halos the size of dwarf galaxies \citep{ghi2001}.

We present a dynamical study of the Fornax Cluster, a nearby poor, but
relatively dense cluster.  Our sample includes a significant number of
dwarf galaxies which form a distinct population kinematically.  We
find strong evidence that Fornax has a distinct subcluster containing
a high concentration of star-forming galaxies. For our analysis we
adopt a cluster distance of 20\Mpc\ \citep[][but see discussion
below]{mou2000}.

%%%%%%%%%%%%%%%%%%%%%%%%%%%%%%%%%%%%%%%%%%%%%%%%%%%%%%%%%%%%%%%%%%%%%%%%%%
\section{Galaxy Sample}
\label{sec_sample}

Our analysis is based on a large study \citep{dri2001a} of the Fornax
Cluster made with the FLAIR-II spectrograph on the UK Schmidt
Telescope (UKST).  We obtained redshifts for 516 galaxies with
$\bj\leq 18$ in a $5.8\times5.8$ \degsq\ field centered on the Fornax
Cluster.  Most of the the targets (426 or $\approx30\%$ of all
galaxies in the field with $16.5<\bj<18.0$) were compact galaxies
(scale lengths $\leq 4$ arcseconds) selected from a digitized UKST
plate to search for new compact cluster members.  The spectra have a
resolution of 13\AA\ (velocity uncertainty $\Delta v=50\kms$).
Eighteen were confirmed as cluster members, nine being new compact
cluster dwarfs \citep{dri1998}.  We also obtained 5.3\AA\ resolution
spectra ($\Delta v=30\kms$) for a sample of 90 galaxies listed as
probable cluster members in the Fornax Cluster Catalog
\citep[FCC]{fer1989}; this is 54\% of the FCC members with $\bj<18$.
All 90 were confirmed as cluster members, giving a total sample of
108.  The 53 galaxies with $\bj<15.5$ ($M_B<-16$) we classify as
giants and the remainder (55) as dwarfs.  We also spectroscopically
classify each galaxy as late-type or early-type; the 42 galaxies with
H$\alpha$ emission equivalent width greater than 1\AA\ (a 3$\sigma$
detection) were classified as late and the remainder (66) as early.
Of the 43 dwarf galaxies with early-type morphological classifications
in the FCC, 11 had enough $H\alpha$ emission to be reclassified by
this criterion, so that $36\pm8\%$ of the dwarfs are spectroscopic
late-types compared to 18\%\ based on morphological classifications.
The implications of this observation on the evolution of star
formation in dwarf galaxies are discussed by \citet{dri2001a}.

\section{Evidence for substructure}

The Fornax Cluster has previously been noted for its relatively smooth
Gaussian velocity distribution \citep{mad1999}.  Using the W test on
our data for Fornax we find that the total sample of 108 galaxies has
a velocity distribution that is marginally non-Gaussian (at the 91\%
confidence level), but the late/early and giant/dwarf subsamples are
all consistent with Gaussians at the 2-$\sigma$ level
(Fig.~\ref{fig_distribution}a,b).  The total sample has a mean
velocity of $1493\pm36\kms$ and a velocity dispersion of
$374\pm26\kms$, in agreement with other work \citep[see][]{dri2000}.
Using the t- and F-tests, we find no significant differences in the
mean velocities of the different subsamples or the velocity
dispersions of the early ($356\pm31\kms$) and late type samples
($405\pm45\kms$).  There is, however, a significant difference (at the
98\%\ level) in the velocity dispersions of the dwarfs
($429\pm41\kms$) and the giants ($308\pm30\kms$).  This was suspected
by \citet{hel1994}, but their sample was too small for a significant
result.  We discuss the implications of this observation in
Section~\ref{sec_discussion} below.

We applied the ``$\kappa$-test'' of \citet{col1996} to the total
sample to test for velocity substructure.  This test takes groups of
the $N$ nearest neighbors to each galaxy and compares the velocity
distribution of the group to that of the whole cluster.  The test
revealed substructure for the $N=4$ case at the 96\%\ significance
level, notably for a group to the southwest of the cluster center.
Fig.~\ref{fig_velocity} shows an adaptively-smoothed projection of the
galaxy number density onto a plane of velocity against distance along
a diagonal vector running through the cluster center from South West
to North East, chosen to isolate the group in the SW corner of the
cluster.  We used the KMM mixture modeling algorithm to locate the
substructure.  This algorithm searches for the maximum-likelihood
partition of a cluster into a specified number of subclusters
\citep[see][]{col1996}.  We used a coordinate system based on offsets
in projected separation on the sky measured in arc minutes from the
position of NGC~1399 ($\alpha=$03:38:29.0 $\delta=-$35:27:01 J2000).
The KMM analysis identified a very robust partition of the sample into
a 92-member cluster (``Fornax-main'') centered at ($3',18',1478\kms$),
with $\sigma_v=370\kms$ and a 16-member subcluster (``Fornax-SW'') at
($-168',-72',1583\kms$) with $\sigma_v=377\kms$.  Fornax-main is
centered near NGC~1399 while Fornax-SW is dominated by NGC~1316
(Fornax~A); see Fig.~\ref{fig_subcluster}.  The estimated correct
allocation rate was 99\%\ and the partition was preferred over a
single distribution at the 99\%\ confidence level.  A further
partition of Fornax-main identified a high-velocity clump (including
the probably infalling NGC~1404 and associated objects, see below)
projected on the cluster center, but this partition was not
significantly better than a single distribution.

The radial distribution of the 92 galaxies in the Fornax-main cluster
alone is shown in Fig.~\ref{fig_distribution}(c,d).  Even on removing
the spiral-rich Fornax-SW subcluster, the remaining 29 late-type
galaxies are much more extended than the 63 early-types (at a KS
significance level of 99\%).  In addition, the Fornax-main sample is
no longer well-described by a King profile \citep{fer1989}, excluded
at a KS significance level of 90\%.  Comparison of the dwarf and giant
distributions in Fornax-main (Fig.~\ref{fig_distribution}(d)) shows
that the dwarfs are more spatially extended at a KS significance level
of 94\%.

\section{Cluster Mass Profile}
\label{sec_dynamics}

The distributions of galaxy velocities against projected radius from
cluster centers are predicted to exhibit caustic structures due to
the overlapping of shells of infalling galaxies \citep{reg1989} but
this has previously been detected only in the Coma cluster
\citep{gel1999}.  We plot this distribution for Fornax in
Fig.~\ref{fig_caustic}.  A caustic-like structure is strongly suggested
when we consider just the early-type galaxies (circles).  If real,
this is the first detection of caustic structure in such a poor
cluster, but the numbers of galaxies are too low to accurately
determine the location of the caustics.  For comparison we
plot one caustic curve for a model \citep{pra1994} with $\Omega_0 =
0.3, \sigma_v = 374\kms$ and a turn-around radius of 2.4\Mpc,
corresponding to a cluster mass of $9\times10^{13}\msun$ and a virial
radius of 0.7\Mpc.

Although the caustics are not well-defined, we can use the method of
\citet{dia1999} to measure the cluster mass as it provides a robust
estimate without making assumptions of dynamical equilibrium.
Briefly, the method works by defining caustic-like curves of the
velocity amplitude ${\cal A}(r)$ as shown in Fig.~\ref{fig_caustic}.
The amplitude is defined by the points where the galaxy density falls
below a threshold $\kappa$, defined to minimize the difference between
the escape velocities calculated from ${\cal A}(r)$ and estimated from
the central velocity dispersion.  The amplitude profile is then
integrated to yield the mass profile $GM(<r) = { 1 \over 2} \int_0^r
{\cal A}^2(x)dx$, also shown in Fig.~\ref{fig_caustic}.  At small
radii the resulting mass profile agrees well with independent
estimates from X-ray measurements at the center of the cluster.  The
total enclosed mass at a radius of 1.4\Mpc\ is
$(7\pm2)\times10^{13}\msun$.  Integrating the blue light from all
galaxies out to the same radius we get $L_{1.4}=2\times10^{11}\lsun$
and a mass-to-light ratio of $300\pm100$ $\msun/\lsun$, in the normal
range for clusters \citep{bah2000}.  Limiting the analysis to the 92
galaxies of the main subcluster we obtain a slightly smaller mass of
$(5\pm2)\times10^{13}\msun$.  We also estimated the cluster mass by
applying the virial theorem to the early-type galaxies alone
(Fig.~\ref{fig_caustic}).  We used the mean of the projected-mass
estimators for radial and isotropic orbits \citep[see][]{hei1985}.
Interestingly, the projected mass estimator based on the early-type
galaxies alone converges to a mass of $9\times10^{13}\msun$ consistent
with that found using the \citeauthor{dia1999} method applied to all
galaxies in the cluster. We also applied the virial (projected mass)
estimator to the SW subcluster, obtaining a mass of
$6\times10^{13}\msun$, but this is an upper limit as this subcluster
is presumably not virialized.  Alternatively, if it had the same
mass-to-light ratio as the main cluster, its mass would be
$2\times10^{13}\msun$.

%%%%%%%%%%%%%%%%%%%%%%%%%%%%%%%%%%%%%%%%%%%%%%%%%%%%%%%%%%%%%%%%%%%%%%%
\section{Discussion}
\label{sec_discussion}

As the difference in velocity dispersions between the dwarfs
($429\pm41\kms$) and the giants ($308\pm30\kms$) is consistent with
the expected ratio of $\sqrt 2$:1 for infalling and virialized
populations \citep{col1996}, we conclude that the dwarf population is
dominated by infalling objects whereas the giants are virialized.  The
more extended spatial distribution of the dwarfs supports this
conclusion.  This is the first detection of such separation of cluster
components by mass, and it is not consistent with numerical
simulations of cluster formation which do not show any mass
segregation in the velocity dispersions \citep{ghi2001}.  An
alternative explanation is that dynamical relaxation (e.g.\ by
dynamical friction of the massive galaxies) could have caused the mass
segregation.  The timescale for this is $\sim 2$ Gyr for the Fornax
Cluster.  Even assuming the dwarfs are only 5 times less massive than
the giants (for a mass-to-light ratio ten times that of the giants),
relaxation would give a dwarf velocity dispersion $\sqrt 5$ times that
of the giants.  We cannot rule out partial relaxation giving the
observed ratio, but this requires fine tuning of the mass segregation
compared to the picture of infalling/virialized samples which
naturally predicts the $\sqrt 2$:1 ratio observed here and in the
Virgo and Coma clusters.

The Fornax-SW subcluster has a recession velocity $(105\pm102)\kms$
greater than the Fornax-main cluster.  We attempted to constrain the
physical motion of the subcluster using a two-body model as in
\citet{col1996}.  The small relative motion of the two components gave
a range of possible solutions, both incoming and outgoing.  All the
solutions were bound.  Assuming all projection angles are equally
probable, the most likely solutions have Fornax-SW infalling at
velocities of 100--500\kms\ at radii of 3.5--1.1\Mpc.  An infalling
solution is supported by individual galaxy properties: Fornax-SW is a
site of intense star formation, containing 15\% of all galaxies in the
total Fornax sample, but 31\% of the star-forming galaxies.  This
subcluster contains two of the four cluster galaxies with
exceptionally high star-formation rates, NGC~1341 and FCC~33
(Figs.~\ref{fig_velocity} and \ref{fig_subcluster}) and
a large amount of neutral hydrogen \citep{put1998},
unlikely to have survived a passage through the cluster core.
Furthermore, observations of Fornax~A in the subcluster
\citep{eke1983} indicate that it has a projected velocity of
approximately 80\kms\ northwards, consistent with infall.  Similarly,
the subgroup seen in projection on the cluster core (seen in
Fig.~\ref{fig_velocity} at 450\kms) is probably real and infalling,
although it was not robustly identified by the KMM algorithm.  This
group contains NGC~1404, which has a distorted X-ray envelope
indicative of infall \citep{jon1997}, and the irregular galaxy NGC
1427A, which shows signs of being disrupted by its first passage
through the cluster \citep*{cha2000}.  The morphologies of both these
galaxies are indicative of material blown off behind them as they move
towards the cluster center.

The substructure that we have identified in Fornax bears on
the determination of its Cepheid distance.  There are now three
Cepheid distances to spirals in the Fornax region: NGC~1365 at
18.6\Mpc\ \citep{mad1999}, NGC~1326A at 18.7\Mpc\ \citep{pro1999}, and
NGC~1425 at 22.2\Mpc\ \citep{mou2000}.  Mould et al.\   suggest that
the mean of these ($\approx20\Mpc$) be adopted, but this may still not
yield an accurate cluster distance.  Though seen in projection only
70\arcmin\ ($\sim 0.4\Mpc$) from the cluster core and with a similar
velocity, NGC~1365 (see Fig.~\ref{fig_velocity}) may be situated near
the turn-around radius ($\sim$2\Mpc\ out or $\sim$10\% in
distance), given the tendency of late-type galaxies to avoid the
cluster core.  If NGC~1365 were in the cluster core, it might be
expected to show morphological peculiarities like spirals in Virgo,
Coma, and other clusters \citep[see][]{bra2000},
but it is symmetric at wavelengths from optical to 20\cm
\citep{lin1999}.  With an identical Cepheid distance, and as a member
of the infalling Fornax-SW subgroup (Fig.~\ref{fig_subcluster}),
NGC~1326A is also a doubtful indicator of the distance to the
Fornax core.  NGC~1425 is perhaps a better gauge
of the cluster distance, but it is 5\fdg6 ($\sim 2\Mpc$) or $\sim 10$
core radii removed from the cluster center; it is not even clear
whether NGC~1425 belongs to Fornax or the nearby Eridanus cluster
(Mould et al.).  While a simpler system than the Virgo cluster,
Fornax nevertheless presents its own difficulties for an accurate
distance measurement using Cepheids; a secure result awaits
observations of additional spirals unambiguously residing in the
cluster core.

%%%%%%%%%%%%%%%%%%%%%%%%%%%%%%%%%%%%%%%%%%%%%%%%%%%%%%%%%%%%%%%%%%%%%%%%%%
\acknowledgments

We are very grateful to B.\  Holman, M.\  Brown and the staff
of the UKST for observing assistance and N.\  Ryan for
the caustic calculations. We thank M.\  Geller, B.\  Moore,
P.\  Thomas and
R.\  Webster for helpful discussions.  Part of this work was done
at the Institute of Geophysics and Planetary Physics, under the
auspices of the U.S.\ Department of Energy by Lawrence Livermore
National Laboratory (contract W-7405-Eng-48).  This material is
based upon work supported by the National Science Foundation
(grant 9970884).

\clearpage

{
\epsscale{0.525} \plotone{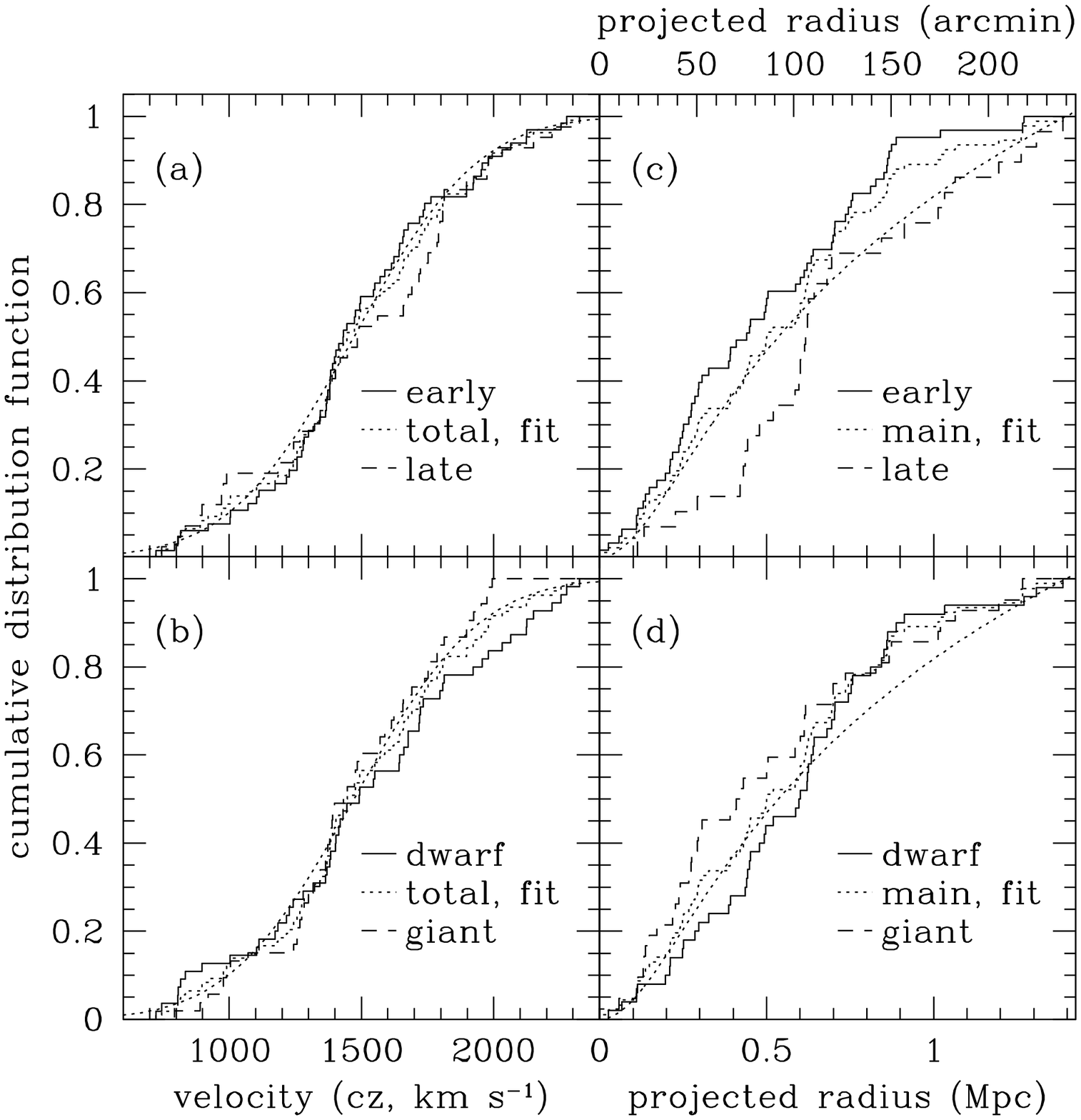}
\figcaption[drinkwater_fig1.eps]{Cumulative distribution functions of
Fornax galaxy velocities (a,b) for the total sample and projected radii
(c,d) for the Fornax-main group only. In the top panels the total samples
are compared to the early- and late-type subsamples and the bottom
panels they are compared to the giant and dwarf
subsamples. Also shown is the Gaussian fit to velocities of the total
sample (a,b) and the King model fitted to the radial distribution of
the whole cluster \citep{fer1989} (c,d).
\label{fig_distribution}}
\smallskip
}

\clearpage

{
\epsscale{0.5} \plotone{drinkwater_fig2.eps}
\figcaption[drinkwater_fig2.eps]{
Adaptively-smoothed plot of the number
density distribution of Fornax galaxies projected onto a plane of
velocity against distance along a vector
running South West to North East.
The five brightest cluster galaxies are marked
by circles for early-types (NGC~1380, NGC~1404 and NGC~1399 from top to
bottom) and crosses for late-types (Fornax A and NGC~1365 from top to
bottom).  Also indicated by asterisks in squares are the four
late-type galaxies with very strong star-formation activity (NGC~1341,
FCC 35, FCCB 2144 and FCC 322, top to bottom).
\label{fig_velocity}}
\smallskip
}

\clearpage

{
\epsscale{0.5} \plotone{drinkwater_fig3a.eps} 
\epsscale{0.523} \plotone{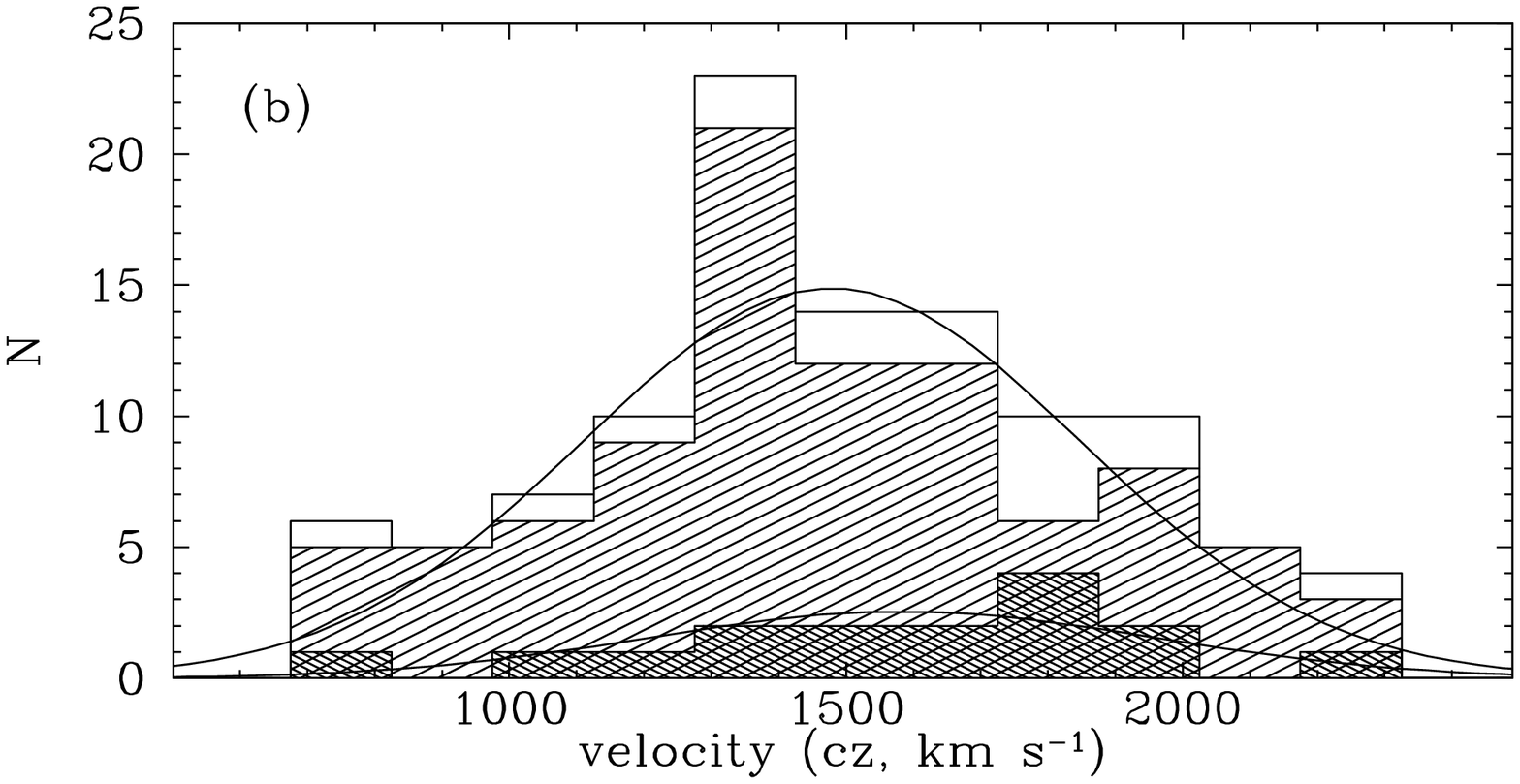}
\figcaption[drinkwater_fig3a.eps]{
Partition of the Fornax system into two subclusters by the KMM
algorithm. (a) Adaptively-smoothed plot of the number density
distribution of Fornax galaxies on the sky (grey
scale). The positions of cluster members are shown by circles (early-type
galaxies) and crosses/asterisks (late-type galaxies). The
asterisks indicate galaxies with high star formation rates. The symbol sizes
are proportional to luminosity. The dashed ellipses are the $2\sigma$
limits of the subclusters. (b) Velocity histograms of the total sample
(unshaded), the Fornax-main cluster (light shading) and the Fornax-SW
subcluster (heavy shading) with the fitted Gaussians overlaid.
\label{fig_subcluster}}
\smallskip
}

{
\epsscale{0.535} \plotone{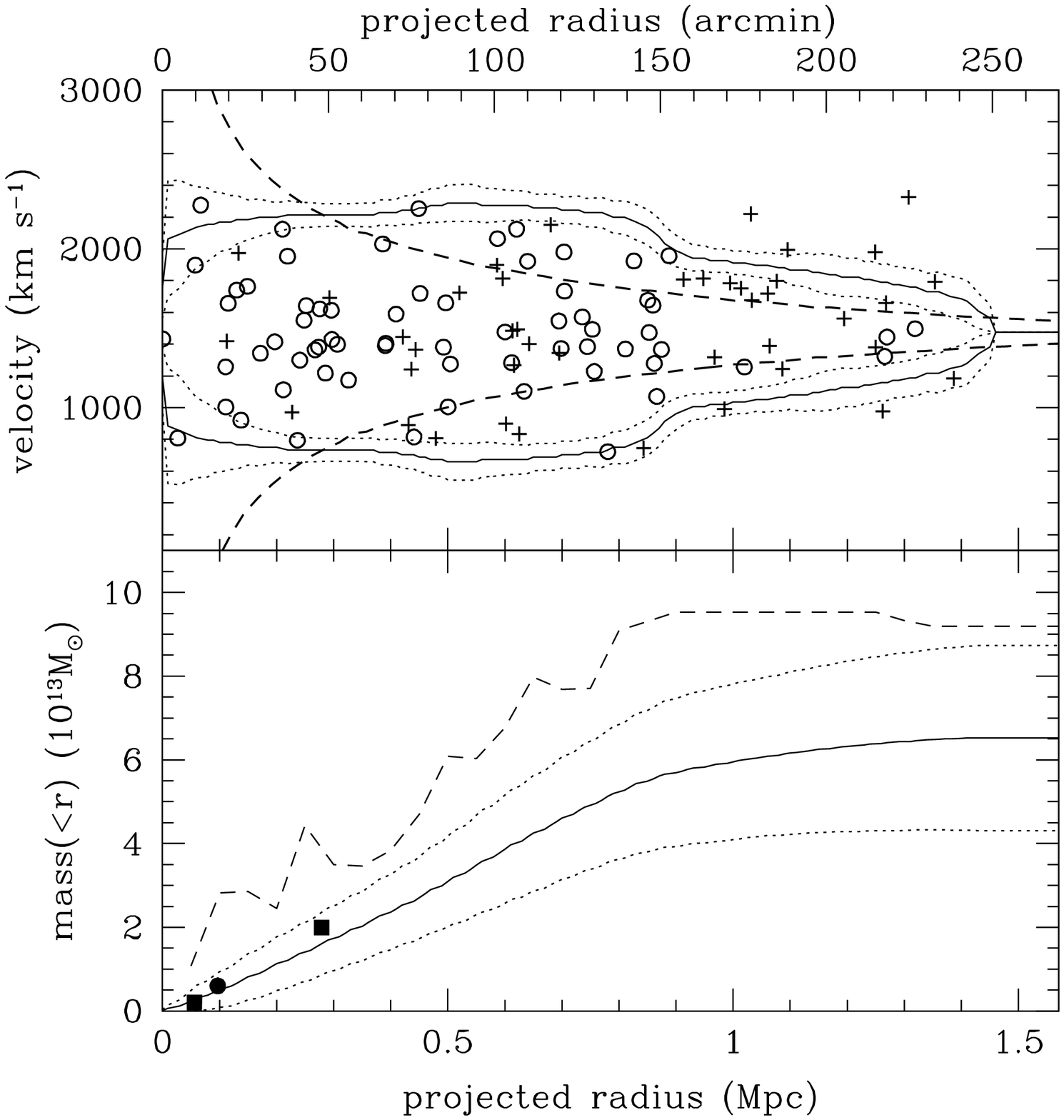}
\figcaption[drinkwater_fig4.eps]{Top: radial velocities of Fornax galaxies
plotted against projected radius from the cluster center. Early type
galaxies are circles and late-types are crosses. The solid and dotted
lines indicate the locus of the velocity amplitude ${\cal A}(x)$ and
its uncertainties discussed in the text. The dashed lines indicate the
locus of a caustic curve for a  model with $\Omega_0 = 0.3,
\sigma_v = 374\kms, r_{turn} = 2.4\Mpc$.  Bottom: Integrated mass
profile of the Fornax Cluster derived by integrating the velocity
amplitude profiles (solid and dotted lines),
from the projected mass virial estimator
(dashed line) and from X-ray observations by \citet{ike1996}
(squares) and \citet{jon1997} (circle).
\label{fig_caustic}}
\smallskip
}

\end{document}